\documentclass[twocolumn,aps,showpacs,floatfix,prc]{revtex4-2}
\usepackage[dvips]{epsfig}
\usepackage{amsmath,bm}
\usepackage{xcolor}[dvips]

\begin{document}

\title{Magnetic field evolution of X-ray emitting radio-quiet pulsars}

\author{Debasis Atta$^{1\dagger}$, Vinay Singh$^{2\S}$ and D. N. Basu$^{3\S}$  }

\affiliation{$^{\dagger}$Department of Higher Education, Government of West Bengal, Bikash Bhavan, Saltlake, Kolkata 700091, India}

\affiliation{$^{\S}$Variable Energy Cyclotron Centre, 1/AF Bidhan Nagar, Kolkata 700064, India }

\email[E-mail 1: ]{debasisa906@gmail.com}
\email[E-mail 2: ]{vsingh@vecc.gov.in}
\email[E-mail 3: ]{dnb@vecc.gov.in}

\date{\today }

\begin{abstract}

    The intense magnetic fields present in neutron stars are closely linked to their observed temperature and spectral characteristics, timing properties, including spin period and its derivatives. Therefore, a comprehensive theoretical analysis of magnetic field evolution is essential for understanding how the strength of the magnetic field change over time. The decay rate of magnetic field in isolated, non-accreting neutron stars can be assessed by evaluating the second derivative of the spin frequency. Another method to estimate this rate involves monitoring an increase in thermal emission beyond what is expected from standard cooling processes, assuming no additional heating mechanisms are present. Our findings indicate that for X-ray emitting isolated neutron stars, the evolution rate of spin period derivative aligns with the dissipation rate of magnetic energy from the dipolar field, provided that a substantial portion of the released energy is emitted as X-rays. The time scale of magnetic field decay is found to be much shorter than typical age of radio pulsars.
   
\vspace{0.2cm}
\noindent
{\it Keywords}: Pulsars; Magnetized neutron star; Magnetic field decay; RX J0420.0-5022; Magnificent Seven.
\end{abstract}

\pacs{95.30.-k; 95.85.Nv; 95.85.Sz; 97.60.Gb; 98.70.Qy}   
\maketitle

\noindent
\section{Introduction}
\label{section1}

    Presently, there are seven nearby radio-quiet isolated neutron stars identified in ROSAT data, characterized by their thermal X-ray spectra. These stars display remarkably similar properties, and despite extensive searches, their population has remained unchanged since 2001, leading to their designation as `The Magnificent Seven' (M7) referring to this collection of isolated young cooling neutron stars located between 120 and 500 parsecs from Earth. Among these, five stars exhibit pulsations in their X-ray flux, with periods ranging from 3.4 seconds to 11.4 seconds. Observations from XMM-Newton have uncovered broad absorption lines in the X-ray spectra, which are interpreted as cyclotron resonance absorption lines caused by protons or heavy ions, and/or atomic transitions that have been shifted to X-ray energies due to the intense magnetic fields, approximately 10$^{13}$ G. Recent XMM-Newton observations suggest that the X-ray spectra are more complex, featuring multiple absorption lines. Pulse-phase spectroscopy of the most extensively studied pulsars, RXJ0720.4–3125 and RBS 1223, reveals variations in the derived emission temperature and the depth of absorption lines as a function of pulse phase. Additionally, RXJ0720.4–3125 exhibits long-term spectral variations which are thought to result from the free precession of the neutron star. Modeling pulse profile of RXJ0720.4–3125 provides insights into surface temperature distribution of these neutron stars, indicating the presence of hot polar caps that vary in temperature and size, and are likely not situated in antipodal positions. 
    
    These pulsars are also known as X-ray dim Isolated Neutron Stars or simply XINS which thermally emit X-rays and optics with no detected radio emission \cite{Bo2023,Po2023}. These objects exhibit stable behavior rather than transient characteristics. Their thermal X-ray spectra lack substantial evidence of a power-law tail. Such attributes distinguish M7 from Galactic magnetars that possess comparable spin periods, significantly constraining the potential influence of the magnetosphere on X-ray emission and timing irregularities associated with M7. These sources are characterized by strong dipolar magnetic fields, typically exceeding $10^{13}$~G. The surface temperatures of M7 are unexpectedly elevated for their ages, suggesting the absence of additional heating mechanisms \cite{Po2018}. Furthermore, their surface thermal luminosities surpass their spin-down luminosities. Consequently, it is frequently posited that they are thermally energized by magnetic field decay and are considered to be the evolutionary successors of magnetars e.g., \cite{Po2010,Vi2013}.

    Recently, Bogdanov and Ho \cite{Bo2024} reported the inaugural measurement of the second derivative of the spin period, $\ddot P=-4.1\times 10^{-25}$~s~s$^{-2}$, for one of the M7 sources, specifically RX J0720.4-3125. This measurement yields the braking index $n =\nu \ddot \nu /\dot \nu^2 \equiv 2 - \ddot P P / (\dot P)^2 \approx 675$ for the pulsar. This value significantly exceeds the expected braking index of $n=3$, which is associated with magnetic dipole braking in a vacuum. Bogdanov and Ho \cite{Bo2024} suggest that this elevated value may be attributed to irregularities in the spin behavior. 
    
    In the present work, we explore the rapid decay of magnetic field in M7 leading to large braking indices (far from its classical value of 3) observed in these XINS which are quite similar to isolated radio pulsars \cite{Ig2020}. We also show that the external dipolar magnetic field of M7 objects is currently diminishing at a rate of about $10^4-10^5$ years. 

\noindent
\section{Time evolution of magnetic field}
\label{section2}

    We present an estimation of the instantaneous decay of magnetic field, derived from the observed braking index. Subsequently, we assess the necessary timescale for magnetic field decay that would facilitate thermal X-ray emission through the Ohmic heating of the crust. Finally, we will compare these timescales.

\vspace{-0.5cm}  
\subsection{Braking index correlation to decay timescale }

    The magneto-rotational evolution of neutron stars (NSs) can be characterized using a simplified form of the magneto-dipole equation for analysis \cite{Ph2014}:

\begin{equation}
    I\nu \dot \nu \propto B^2 \nu^4
\end{equation}
where $\nu$ is the frequency, $B$ is the surface magnetic field of a NS and $I$ its moment of inertia. 

    Introducing the braking index serves as a useful characteristic for understanding the evolution of spin which is defined $n=\nu \ddot \nu /\dot \nu^2$. In the case of spin-down under a constant magnetic field and other parameters of the neutron star, Eq.~(1) yields a value of $n=3$. Conversely, a diminishing magnetic field leads to $n>3$.

    When the magnetic field exhibits a non-zero first derivative while maintaining constant parameters such as the moment of inertia and magnetic inclination, a straightforward algebraic manipulation results in:

\begin{equation}
    \frac{\dot \nu^2}{\nu^4}(n-3)\propto B\dot B.
    \label{tau}
\end{equation}
A significant positive braking index indicates a negative derivative of the magnetic field, which results in a reduction of the magnetic field strength.

    To obtain an estimate for the timescale of magnetic field decay, it is canonical to postulate that the field diminishes exponentially as

\begin{equation}
   B=B_0 \exp{\left(-t/\tau\right)}. 
\end{equation}
In this context, $B_0$ represents the initial field, while $\tau$ denotes a specific characteristic time scale associated with decay. It is important to note that $\tau$ may be influenced by $B$ or other parameters; however, our focus is on the instantaneous value of $\tau$. This presumption yields

\begin{equation}
    \dot B= - \frac{B}{\tau}.
    \label{bdot}
\end{equation}

    Using Eqs.~(\ref{tau}, \ref{bdot}) we obtain
    
\begin{equation}
    \tau = - \frac{2 \nu}{(n-3)\dot \nu}.
    \label{taubn}
\end{equation}

\vspace{-0.5cm}
\subsection{Ohmic heating of the crust}

    Let us consider the scenario in which all magnetic energy released during decay is converted into Ohmic heating of the crust, subsequently resulting in the emission of X-rays that contribute to the luminosity $L_\mathrm{X}$. This representation is somewhat simplified, as a portion of the heat may be conducted inward to the core, where it can be emitted as neutrinos \cite{Ka2014}. In cases of relatively low energy release, as anticipated for M7 sources, a significant fraction of the energy is emitted from the surface, particularly when the magnetic energy is released at shallow depths within the crust and direct URCA processes remain inactive in the core. The presence of superfluidity within the crust enhances the surface temperature for a given amount of energy release and depth \cite{Ka2017}. Furthermore, there may be additional luminosity contributions from residual heat emission from the surface.

    The energy associated with the magnetic field can be approximated as
    
\begin{equation}
   E_\mathrm{mag}=\left(\frac{B^2}{8\pi} \right) \left(\frac{4}{3}\pi R^3 \right) = \frac{B^2R^3}{6} 
\end{equation}
where $R$ is the NS radius. 

    Assuming that the magnetic field is localized within the crust, which has a volume of approximately $V\approx 4 \pi h R^2$ and a thickness $h$ of about one-third of a km, the estimate for the total magnetic energy is diminished by an order of magnitude. Additionally, we do not account for contributions from non-dipolar (and non-poloidal) field components. It is important to mention here that for RX J0720.4-3125, there is evidence indicating a significant non-dipolar external magnetic field \cite{Bo2017}. The subsequent release of energy is as follows

\begin{equation}
    \dot E_\mathrm{mag}=\frac{B\dot B R^3}{3}.
\end{equation}

    Assuming that $\dot E_\mathrm{mag}\equiv L_\mathrm{X}$ and using Eq.~(\ref{bdot}) yields
    
\begin{equation}
L_\mathrm{X} = \frac{B^2 R^3}{3\tau}    
\label{eq:lum}
\end{equation}
which in turn provides estimate for time scale as

\begin{equation}
\tau=\frac{B^2R^3}{3L_\mathrm{X}}
\label{taub}
\end{equation}

    It is important to highlight that the magnetic field associated with M7 objects can be assessed through two distinct approaches. The first method involves utilizing the spin period and its derivative, while the second method relies on the characteristics of the proton cyclotron line. Both approaches yield consistent results \cite{Ha2007}. In the following estimates, we will employ the magnetic field values obtained from the spin-down rate using the magnetic dipole model \cite{Ja2007} of pulsars as
    
\begin{equation}
 B= 3.2\times 10^{19} \Big(\frac{-\dot\nu}{\nu^3}\Big)^{1/2} {\rm G}.
    \label{bdip}
\end{equation}     

\noindent
\section{ The Magnificent Seven }
\label{section3}

    We present a systematic X-ray pulse timing analysis of the six members of the so-called `Magnificent Seven' nearby thermally emitting isolated neutron stars (XINS) with detected pulsations.

    The assessment of the second derivative of the spin period for a neutron star, which is believed to be driven by the decay of its magnetic field, alongside an independent measurement of the magnetic field obtained through spectral data, presents a distinctive opportunity to investigate whether its spin evolution aligns with its thermal surface emission.

    RX J0720.4-3125 exhibits a magnetic field strength of $B=2.5\times 10^{13}$G and a frequency derivative of $\dot \nu = -1.02\times 10^{-15}$. This magnetic field value is calculated using the magneto-dipole formula and aligns well with the approximate value of $B\approx 5\times 10^{13}$G, which was derived from spectral features interpreted as resulting from cyclotron resonance scattering of protons in the magnetic field \cite{Ha2004}. From Eq.~(\ref{taubn}), using $n=675$ \cite{Bo2024} we determine that $\tau=1.1\times 10^4 {\rm yrs}$. Utilizing this value to estimate the expected luminosity through Eq.~(\ref{eq:lum}), we find that the value of $L_\mathrm{X}$ is, approximately, $6\times 10^{32}$~erg~s$^{-1}$. Alternatively, if the magnetic field is confined to the crust, the luminosity is approximately $L_\mathrm{X}\approx 0.6\times 10^{32}$~erg~s$^{-1}$.

    It is noteworthy that this value is remarkably similar to the luminosity of RX J0720.4-3125, which is approximately $(1.1-3.2)\times 10^{32}$~erg~s$^{-1}$ \cite{Po2020}. This indicates that the time scale obtained from the evolution of the spin period aligns well with the scale $\tau$ as presented in Eq.~(\ref{taub}). Furthermore, considering that our estimate utilizes the total volume of the neutron star and acknowledges that a portion of the energy may be released as neutrinos, the correlation remains strong, as the observed luminosity is less than that calculated from Eq.~(\ref{eq:lum}).

    The time scale obtained for RX J0720.4-3125 is significantly shorter than its kinematic age of approximately 0.4 to 0.5 million years \cite{Te2010}. Furthermore, this time scale does not align with the Hall time scale, which is given by $\tau_\mathrm{Hall}\sim(10^{18}-10^{19}) B^{-1}$ years \cite{Ag2008}, when we apply the external dipolar field values determined for the M7 objects.
     
    The numerical simulations concerning the magneto-thermal evolution of neutron stars indicate that M7 sources exhibit significant braking indices, as evidenced by their nearly vertical trajectories in the $P$~--~$\dot P$ diagram presented in Fig.-10 by Vigan\'o et al. \cite{Vi2013}. By extracting numerical values for $\dot P$ and ages from the track corresponding to an initial magnetic field of $B = 3\times 10^{14}$~G in that diagram, one can estimate $\ddot P\approx 4\times 10^{-26}$~s s$^{-2}$, which corresponds to a braking index of approximately $n\approx 68$. To account for values of $n\sim 221$, it appears necessary to postulate a period of accelerated magnetic energy dissipation, potentially resulting from some form of instability.
    
    Recently, timing measurements have been acquired for another M7 object, RX J0806.4-4123 \cite{Po2024}. Despite the fact that all other indicators suggest that RX J1605.3+3249 is highly likely to be an X-ray isolated neutron star (XINS), no X-ray pulsations have been observed from this source thus far, including a reported periodicity of 3.4 seconds by Pires et al. \cite{Pi2014}.   
    
\noindent
\section{ Calculations and results }
\label{section4}

\begin{figure}[t]
\vspace{0.0cm}
\eject\centerline{\epsfig{file=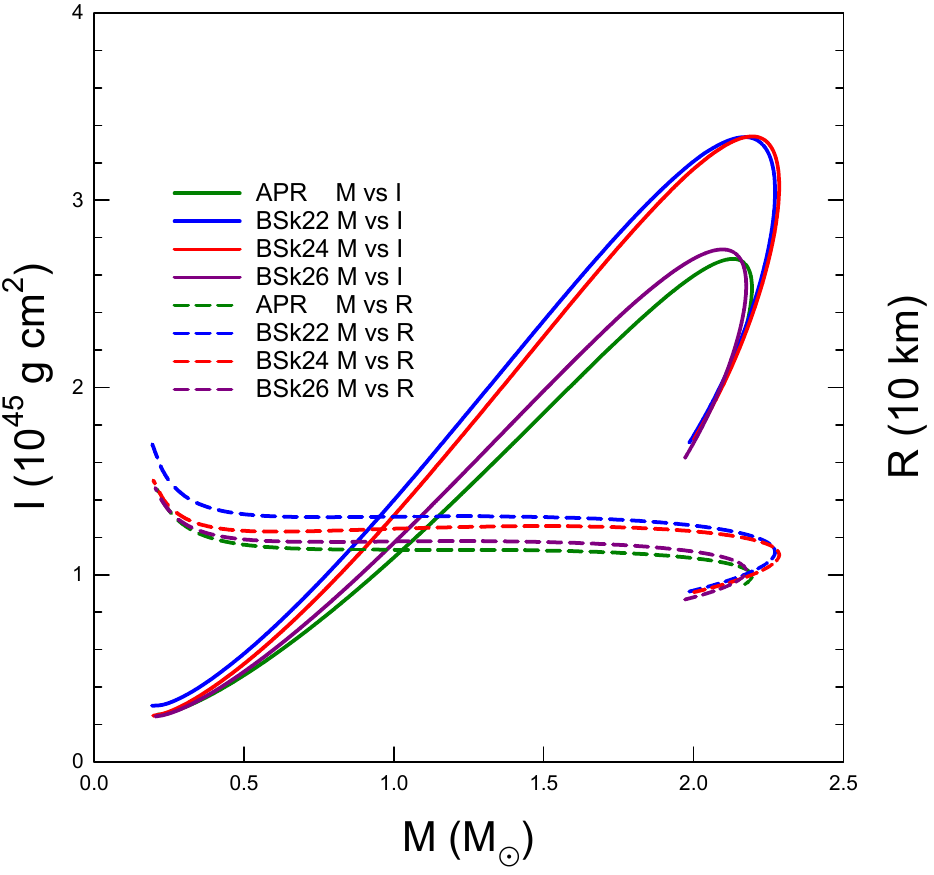,height=8cm,width=8cm}}
\caption{Plots of moment of inertia I in 10$^{45}$ g cm$^2$ unit and radius R in 10 km unit vs mass M in solar mass M$_\odot$ unit for pulsars obtained using APR, BSk22, BSk24 and BSk26 EoSs.} 
\label{fig1}
\vspace{-0.5cm}
\end{figure}

    We present a systematic analysis of the seven members of the so-called `Magnificent Seven' nearby thermally emitting isolated neutron stars (XINS) with detected pulsations utilizing the recent data from \cite{Bo2024}.

\begin{equation}
n=\frac{6\nu L_\mathrm{X}}{B^2 R^3\dot \nu} + 3
\end{equation}
We anticipate that the braking indices for the remaining M7 sources will be in the range of several hundreds. In a similar manner, we can also estimate the values of $\ddot \nu$ as

\begin{equation}
    \ddot \nu= -\frac{6n L_\mathrm{X}\dot \nu}{(n-3)B^2R^3}.
\end{equation}
The characteristic or spin-down age of a pulsar $\tau_c$ given by 

\begin{equation}
    \tau_c = -\frac{\nu}{2\dot \nu}
\end{equation}
is a useful parameter but can not be used as true age, yet provides its upper limit. As $\nu$ is of the order of s$^{-1}$ or Hz while $\dot \nu$ is of the order of $10^{-15}$ Hz s$^{-1}$, its value turns out to be in the range of mega year (Myr). The spin-down luminosity generated by consuming the rotational energy of a pulsar is given by 

\begin{equation}
    L_\mathrm{r} = -4\pi^2 I \nu \dot \nu
\end{equation}
where $I$ is the moment of inertia of the star. In Fig.-\ref{fig1}, the moment of inertia and radius versus mass of pulsars generated \cite{La24} from Equation of States (EoSs) obtained using Akmal-Pandharipande-Ravenhall (APR) \cite{Ak98} and Skyrme effective interactions with Brussels-Montreal parameter sets (BSk22, BSk24, BSk26) \cite{Gor13} have been plotted. It suggests that for model calculations presented here, typical values of $I=10^{45}$ g cm$^2$ and $R=10$ km can be used universally. Should the energy loss be attributed to the magnetic field decay, the ratio $L_\mathrm{r}/L_\mathrm{X}$ is expected to be a few orders of magnitude less than unity with a likely evolutionary relationship with magnetars. 

\vspace{-0.5cm}
\subsection{Predictions for M7 sources}

    The results for the estimated magnetic field $B$, braking index $n$ (rounded to nearest integer), second time derivative $\ddot \nu$ of frequency and time scale $\tau$ for magnetic field decay have been presented in Table-1. The thermal luminosities $L_\mathrm{X}$ of X-ray have been taken from Potekhin et al. \cite{Po2020}. The mean values have always been used in the calculations even if a range is given. The frequency $\nu$ and its time derivative $\dot \nu$ used in the calculations have been obtained from Bogdanov and Ho \cite{Bo2024}. For RX J1605.3+3249 these data have been obtained from Pires et al. \cite{Pi2014}. The values for the magnetic field $B$ has been obtained from $\nu$ and $\dot \nu$ in the framework of magnetic dipole model for pulsars using Eq.~(\ref{bdip}).
 
\begin{table*}
\caption{The estimated braking index and second time derivative of frequency. The luminosities $L_\mathrm{X}$ of thermal X-ray have been taken from Potekhin et al. \cite{Po2020}. Mean values have always been used in the calculations even if a range is given. The frequency $\nu$ and its time derivative $\dot \nu$ have been obtained from Bogdanov and Ho \cite{Bo2024}. For RX J1605.3+3249 these data have been obtained from Pires et al. \cite{Pi2014}.} 

    \centering
    \begin{tabular}{lcccccccl}
    \hline
    Name & $L_\mathrm{X}$&$\nu$ & $\dot \nu$ & $B$ &  $n$ & $\ddot \nu$&$\tau$&$L_\mathrm{r}/L_\mathrm{X}$ \\
         & ($10^{31}$~erg s$^{-1}$) &(Hz) &($10^{-15}$ Hz s$^{-1}$) & ($10^{13}$~G) & & ($10^{-28}$~Hz s$^{-2}$)&(yr)&\\
    \hline
    RX J1308.6+2127  & 33$^{+5}_{-7}$ &0.096969489591$^{(+7)}_{(-7)}$& $-1.05427^{+0.00003}_{-0.00003}$ & 3.4409 & $157$ & 17.9746& 3.7923$\times 10^4$&0.0122 \\

    RX J0420.0-5022  & 0.6$^{+0.2}_{-0.2}$ &0.28960290947$^{(+6)}_{(-5)}$ &
$-2.4393^{+0.0005}_{-0.0005}$ & 1.0141 & $45$ & 9.1555 &1.8117$\times 10^5$&4.6481\\
    
    RX J0720.4-3125  & 19$^{+13}_{-8}$ &0.1191733514$^{(+1)}_{(-2)}$ & $-1.0156^{+0.0004}_{-0.0004}$ & 2.4788 & $221$ & 19.1023 & 3.4182$\times 10^4$&0.0251 \\

    RX J1605.3+3249  & 0.7$-$50 &0.29517123$^{(+140)}_{(-140)}$&$-139.0$ & 7.4395 & $4$ & 2345.6971 & 2.3077$\times 10^5$&6.3896 \\

    RX J0806.4-4123 & 1.6$-$2.5 &0.08794776394$^{(+2)}_{(-2)}$&$-0.0818^{+0.0001}_{-0.0001}$& 1.1097 & $1077$ & 0.8194 & 6.3489$\times 10^4$&0.0139\\
    
    RX J2143.0+0654  & 5$-$17 &0.10606446051$^{(+9)}_{(-14)}$&$-0.4663^{+0.0006}_{-0.0004}$& 2.0005 & $378$ &7.7520 &  3.8453$\times 10^4$&0.0178\\
    
    RX J1856.6-3754  & 5$-$8 &0.14173936875$^{(+2)}_{(-2)}$&$-0.60373^{+0.00008}_{-0.00007}$& 1.4735 & $425$ & 10.9223 & 3.5304$\times 10^4$&0.0520 \\
    \hline
    \end{tabular}
    \vspace{-0.77cm}
    \label{tab:predicted_n}
\end{table*}

    In order to justify present theoretical approach, it is important to mention here that Ertan et al. \cite{Er2014} suggested that the M7 objects are neutron stars encircled by a fallback disk. Specifically, these researchers presented a model that outlines the spin period, period derivative, and corresponding X-ray luminosities. The assessment of the second spin period derivative offers a chance to validate the model proposed by Ertan et al. However, Igoshev et al. \cite{Ig2024} have arrived at an approximate braking index of $n\approx 80$. While this estimate has the correct sign, it is nearly an order of magnitude lower than the actual value reported by Bogdanov and Ho \cite{Bo2024}. In addition to the evident discrepancy between the predictions of the fallback disk model and the measurements of the braking index, there are further reasons to challenge the fallback disk hypothesis. According to \cite{Bo2024}, five of the M7 objects exhibit a consistent spin-down without any signs of significant anomalies in their long-term timing behavior. Furthermore, all accreting sources demonstrate considerable X-ray variability, a characteristic that is notably absent in M7. It is important to highlight that the magnetic field values suggested for the M7 objects by Ertan et al. \cite{Er2014} do not align with the values derived from the spectral features, which are based on the supposition of proton cyclotron resonance lines.
    
    As obvious from Table-I, except for RX J0420.0-5022 and RX J1605.3+3249 the energy loss can be attributed to the decay of magnetic field, since the ratio $L_\mathrm{r}/L_\mathrm{X}$ is about two orders of magnitude less than unity and hence, likely to have evolutionary relationships with the magnetars. Moreover, the characteristic or spin-down ages $\tau_c$, which are of the order of Myr, are a few orders of magnitudes higher than magnetic field decay time $\tau$. For RX J0420.0-5022 and RX J1605.3+3249 $L_\mathrm{r}/L_\mathrm{X}$ are higher than unity and, correspondingly, an order of magnitude higher values of $\tau$.       

\vspace{-0.5cm}
\subsection{Delayed magnetic field decay mechanisms}

    The acceptance of the evidence outlined in the preceding sections prompts a discussion on the potential physical mechanisms that may drive the evolution of magnetic fields over an exceptionally brief time frame of $10^4$ years, following extensive periods of slower evolution lasting hundreds of thousands of years. Current estimates regarding the decay of magnetic fields due to crust resistivity range from 0.1 to 1 million years for the pasta layer in magnetars \cite{Pons2013}, while for typical radio pulsars, the estimates exceed 10 million years \citep{Ig2019}, with some indications suggesting a decay scale of approximately 30 million years \cite{Ig2019}. The Hall time scale for a crust-confined magnetic field is approximately $\tau_\mathrm{Hall}\sim (10^{18}-10^{19}) B^{-1}$ years, applicable to magnetic fields comparable to dipole estimates. These mechanisms governing magnetic field evolution within the crust do not permit a phase of delayed field decay. Consequently, it is essential to explore alternative sources to account for this evolution. 

    The neutron star core has been frequently proposed as a potential location for the delayed evolution of magnetic fields. It has been established for some time that numerous magnetic field configurations within the core exhibit instability, as noted in \cite{La2012} and its associated references. Research has demonstrated that a purely poloidal magnetic field becomes unstable when subjected to the effects of single fluid ambipolar diffusion \citep{Ig2023}. This instability generates new currents of electricity inside the crust, potentially leading to the release of thermal energy over time scales that are comparable to those of ambipolar diffusion. This time scale is influenced by both temperature and the strength of the magnetic field. The process of ambipolar diffusion necessitates a period for the neutron star to cool and for instabilities to develop, which may account for the delayed onset of the decay. The phenomenon of decay attributed to ambipolar diffusion was previously addressed in relation to RX J0720.4-3125 by \cite{Cr2004}, although the authors focused on a different evolutionary time scale characterized by a more gradual decay.

    An alternative explanation may involve a rapid evolution resulting from the core's transition to a superconductor/superfluid state \cite{Gl2011}. Recent research conducted by \cite{Br2024} indicates that the internal magnetic fields within the crust could experience significant amplification during the process of flux expulsion, which would likely result in increased thermal emission and a hastened evolutionary process. The transition to a superconductor is temperature-sensitive, which may account for the delayed onset observed. Nevertheless, comprehensive investigations into the evolution of the magnetic field within the neutron star core are still in the preliminary stages, yielding only a limited number of reliable findings. Consequently, we are unable to definitively establish that the core evolution is the cause of the heightened X-ray luminosity observed in M7.

\vspace{-0.5cm}
\subsection{RX J0420.0-5022 RX, J1605.3+3249 anomalies}
    
    The ensemble of seven thermally emitting isolated neutron stars, identified by ROSAT and referred to as the Magnificent Seven (M7), stands out distinctly within the diverse populations of neutron stars. The phenomenon of crustal heating, attributed to the decay of magnetic fields, along with a potential evolutionary relationship with magnetars, may elucidate the reasons behind the slower rotation rates and elevated thermal luminosities and magnetic field strengths observed in these objects compared to typical rotation-powered pulsars of equivalent age.  
    
    Whereas the case for RX J0420.0-5022 may be termed as anomalous, the unexpected results for RX J1605.3+3249 are consequence of the fact that in spite of all other indicators implying RX J1605.3+3249 very likely to be an X-ray isolated neutron star, no X-ray pulsations have been detected from this source to date. The claim of a 3.4 second periodicity proposed by Pires et al. \cite{Pi2014} has not been corroborated by later, more extensive observations. The third brightest isolated neutron star, RX J1605.3+3249, remains the sole object among the seven that has not yet exhibited a detected periodicity. The spectrum of this source is entirely thermal, devoid of notable magnetospheric emissions, yet it is intricate, showcasing both narrow and broad absorption features. These characteristics may be utilized to impose constraints on the surface component of the magnetic field and the mass-to-radius ratio of the neutron star. 
    
\vspace{-0.5cm}    
\noindent
\section{ Summary and conclusion }
\label{section6}

    We illustrate a correlation between time scales of magnetic field dissipation derived from braking index and those obtained from the X-ray luminosity for the Magnificent Seven objects (M7). The rate at which the magnetic field decays in isolated, non-accreting neutron stars has been evaluated by analyzing the second derivative of the spin frequency together with observing an increase in thermal emission that exceeds the expectations of standard cooling processes, under the assumption that no additional heating mechanisms are involved. In both analyses, we posit that the decay of the magnetic field is the primary mechanism driving the unusual spin evolution and observed emissions from the surface. It is proposed that the measured luminosity of thermal X-ray may serve as an indicator of evolution of the dipole field. Consequently, predictions regarding the braking index and $\ddot \nu$ for other M7 sources have been we offered. It is anticipated that the braking indices will be approximately a few hundred, with $\ddot \nu$ expected to be in the of $10^{-28}-10^{-27}$~Hz~s$^{-2}$. Our results suggest that for isolated neutron stars that emit X-rays, the rate of change in spin period derivative corresponds with the dissipation rate of magnetic energy from dipolar field, assuming a significant fraction of the released energy is radiated as X-rays. Furthermore, the time scale for magnetic field decay is determined to be considerably shorter than the typical lifespan of radio pulsars.    
 
\vspace{-0.5cm}    
\begin{acknowledgments}

    One of the authors (DNB) acknowledges support from Anusandhan National Research Foundation (erstwhile Science and Engineering Research Board), Department of Science and Technology, Government of India, through Grant No.CRG/2021/007333.
 
\end{acknowledgments}

\end{document}